\documentclass[10pt]{article}
\usepackage{amssymb,overcite}
\usepackage{multicol,epsfig}
\begin{document}

\title{Classical nucleation theory for the nucleation
of the solid phase of
spherical particles with a short-ranged attraction}

\author{{\bf Richard P. Sear}\\
~\\
Department of Physics, University of Surrey\\
Guildford, Surrey GU2 5XH, United Kingdom\\
email: r.sear@surrey.ac.uk}

\date{}

\maketitle

\begin{abstract}
Classical nucleation theory is used to estimate the free-energy
barrier to nucleation of the solid phase of particles
interacting via a potential which has a short-ranged attraction.
Due to the high interfacial tension between the fluid and solid
phases, this barrier is very large,
much larger than in hard spheres. It is divergent in the limit
that the range of the attraction tends to zero.
We predict an upper limit on nucleation in good agreement
with the results of experiments on the crystallisation of proteins.
\end{abstract}

\begin{multicols}{2}

\section{Introduction}

It is difficult to crystallise globular proteins yet it is important
to do so as X-ray crystallography can be performed on crystalline
proteins and the full structure of the protein in its native
state determined \cite{durbin96,rosenberger96}.
Knowing the structure of a protein is
crucial to understanding its function.
Proteins are of course vital to all life.
Crystallising a protein is difficult
as there seem to be only a few conditions
under which a protein solution crystallises but many under
which the solution is either stable on all accessible time scales
or useless amorphous aggregates form. This is in sharp contrast
to the simpler system of hard-sphere-like colloids. The
crystallisation of these is routine. Here we will examine the
first stage of crystallisation: nucleation. We will do so using
classical nucleation theory. Although not a quantitative theory, it
has the merit of simplicity and incorporates the essential features
of nucleation. These are that nucleation is a large amplitude, localised,
fluctuation which forms a microscopic crystallite of the crystalline
phase.

Because of their phase diagram \cite{rosenbaum96,note}
it has been suggested \cite{rosenbaum96} that
the interaction between globular protein molecules can be
approximated by a steeply repulsive core, which prevents two
protein molecules interpenetrating, and an attraction with a range
that is much less than the diameter of the protein; the
surfaces of two protein molecules have to be within a tenth or less
of their diameter for there to be a significant attraction. Here we will
study a simple model potential of this type. In fact, the interactions
between protein molecules are complex and include quite specific,
non-spherically symmetric interactions \cite{durbin96},
thus it should be borne in mind that our model potential
is extremely simplified.

Classical nucleation theory requires as input the chemical potential
of the metastable fluid phase from which the crystal forms,
the chemical potential of the crystalline phase and the surface tension.
The last of these three, the surface tension,
often poses a problem as the surface tension is known for only
a very few model potentials. Indeed even for the simplest potential,
the hard-sphere potential, accurate values for the
surface tension have only been calculated recently
\cite{ohnesorge94,kyrlidis95}. However, recently the author
has calculated the surface tension for a potential
consisting of a hard core and a very short-ranged attraction
\cite{sear99}.

In the next section we will define our model
potential and describe its phase behaviour. We will give the
required expressions
for the chemical potential in the fluid and solid phases.
Then in the third section we will obtain the classical nucleation theory
predictions for our model. The fourth section compares our theory with
experiment and the fifth is a conclusion.

\section{Model and bulk phase diagram}

First, we define the hard-sphere+square-well potential \cite{hansen86}.
It is the spherically symmetric pair potential $u(r)$ defined by
\begin{equation}
u(r)=
\left\{
\begin{array}{ll}
\infty & ~~~~~~ r \le \sigma\\
-\epsilon & ~~~~~~ \sigma<r\le \sigma(1+\delta)\\
0 & ~~~~~~ r > \sigma(1+\delta)\\
\end{array}\right. ,
\label{monoss}
\end{equation}
where $\sigma$ is the hard-sphere diameter, and
$r$ is the separation between the centres of the spheres.
The potential depends on one parameter, $\delta$, the ratio of
the range of the attraction to the hard-sphere diameter. We will
always consider this range to be small: $\delta\ll1$.
The first person to consider very short-ranged
attractions was Baxter \cite{baxter68} who considered a potential
with zero range, $\delta=0$, and with a well-depth $\epsilon/kT$
adjusted so that the second virial coefficient \cite{hansen86}
was of order
unity. $T$ is the temperature and $k$ is Boltzmann's constant.
This model is often termed the sticky-sphere model.
Within it the second virial coefficient is used as
a temperature-like variable. However, Stell \cite{stell91}
showed that the sticky-sphere model was pathological, its fluid
phase is unstable at all non-zero densities.
Therefore, we will not consider this model but instead will
follow Stell when we take the
limit $\delta\rightarrow 0$,
thus obtaining his $\beta_0$ model \cite{stell91,hemmer90,borstnik97}.

\begin{figure}
\caption{
The phase diagram of the $\beta_0$ model
\cite{stell91,hemmer90,borstnik97,sear98}. The solid
lines denote the coexisting densities at the fluid-solid and
solid-solid transitions. The density difference between the
coexisting phases at the solid--solid transition is zero on this
scale \cite{bolhuis94}. The letters denote the fluid (F) and solid (S)
phases, and the two-phase region of fluid-solid coexistence
is marked with the number 2.
The dashed lines are tie lines connecting coexisting fluid and
solid phases.}
\label{sspd}
\begin{center}
\epsfig{file=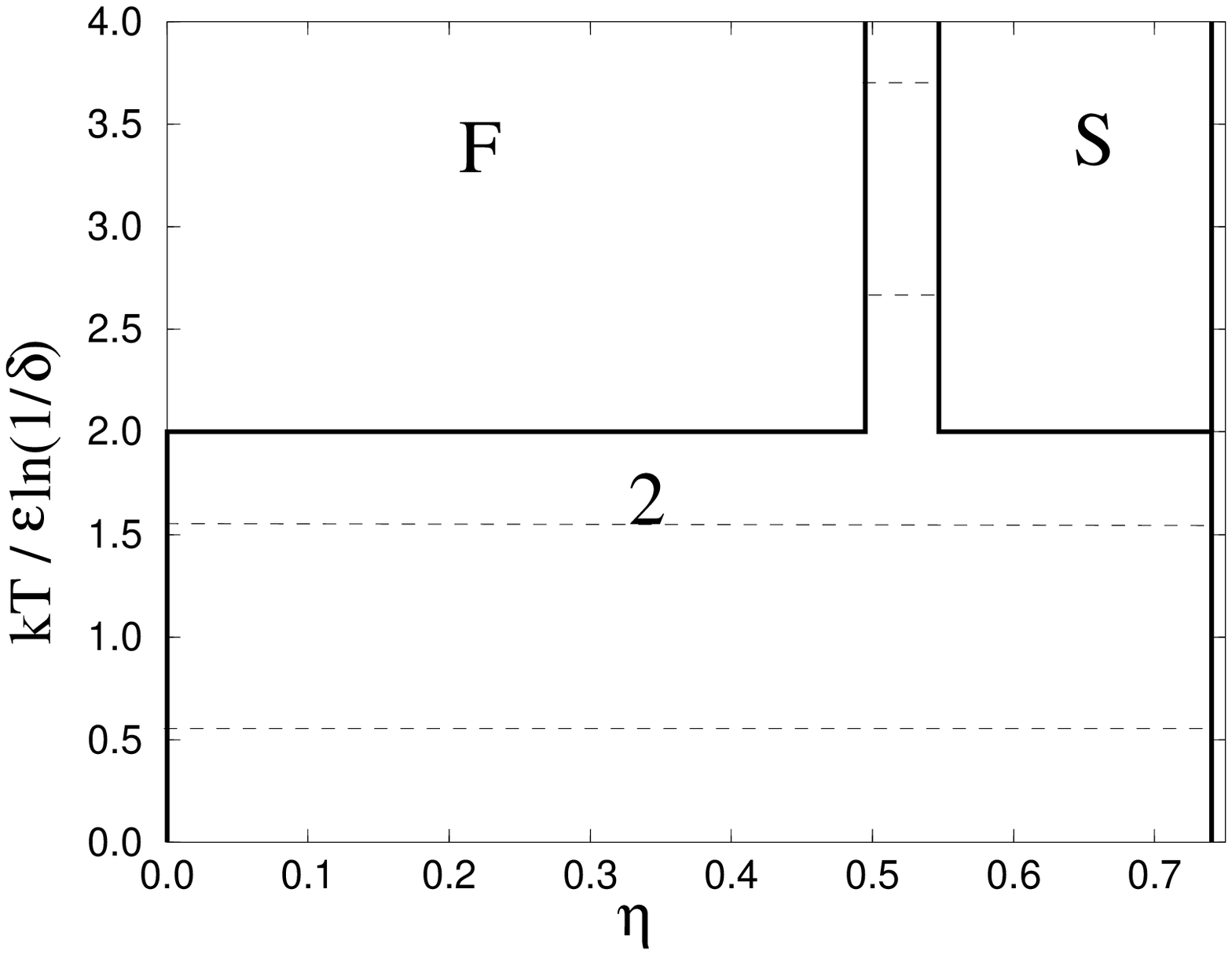,width=3.5in}
\end{center}
\end{figure}

The bulk phase behaviour of the $\beta_0$ model is described in Refs.
\citen{bolhuis94,stell91,hemmer90} and \citen{sear98},
and is displayed in Fig. \ref{sspd}.
If the double limit $\delta\rightarrow0$ and $kT/\epsilon\rightarrow0$
is taken such that $T>T_{coll}$ then
the equilibrium phase behaviour is identical
to that of hard spheres. If $T<T_{coll}$ then
the behaviour is radically different: a close-packed
solid coexists with a fluid phase of zero density.
The temperature $T_{coll}$ is
\cite{hemmer90,stell91,sear98}
\begin{equation}
\frac{kT_{coll}}{\epsilon}=
\frac{2}{\ln(1/\delta)}.
\label{tcoll}
\end{equation}
The close packed density is $2^{1/2}\sigma^{-3}$ which
corresponds to a volume fraction $\eta$ occupied by the hard
cores of $\eta=2^{1/2}\pi/6\simeq0.74$.
At close packing there is an isostructural solid-solid transition
\cite{bolhuis94}.
The bulk phase behaviour of hard spheres with a short-range
square-well attraction, $0.06\ge\delta >0$, is qualitatively similar
\cite{bolhuis94}.
It is essentially Fig. \ref{sspd} but with the corners rounded off and
the horizontal (vertical) lines no longer horizontal (vertical)
but sloping, see Refs.
\citen{bolhuis94,daanoun94} and \citen{hagen94}.
The coexisting densities no longer change
discontinuously but only rapidly near $T_{coll}$.
For $\delta>0.06$ there is no solid--solid
transition but there is no vapour-liquid transition
until considerably larger values of $\delta$.

\subsection{Phase coexistence for nonzero $\delta$}

At temperatures significantly below $T_{coll}$, the density of the fluid
phase which coexists with the solid phase is very low
(zero in the $\delta\rightarrow0$ limit).
So, we will approximate the chemical potential of the fluid
phase by that of an ideal gas.
Around the temperature $T_{coll}$ the density of the fluid phase
which coexists with the solid rapidly increases (in the
$\delta\rightarrow0$ limit it is horizontal, see Fig. \ref{sspd}).
So the accuracy of the approximation will rapidly worsen as
the temperature reaches $T_{coll}$.
The chemical potential, $\mu_f$, of an ideal gas
\begin{equation}
\mu_f=kT\ln\rho,
\label{muf}
\end{equation}
where $\rho$ is the fluid's number density of spheres
times $\sigma^3$ to render it dimensionless.
We have neglected a term $\ln(\Lambda/\sigma^3)$, where
$\Lambda$ is the integral over the momentum degrees of freedom.

At low temperature, the Helmholtz free energy per particle,
$a_s$, of the solid phase can be
calculated by using a cell theory \cite{buehler51},
and assuming that the attractions are strong enough so that
they force the lattice constant to be small enough so that
each sphere is within the range of the attractive interaction
of all twelve of its nearest neighbours \cite{sear98}.
This fixes the energy per particle at $-6\epsilon$ and
the lattice constant at $\sigma(1+c\delta)$,
where $c$ is a constant a little less than one,
see Ref. \citen{sear98}. Then the Helmholtz free energy
per particle $a_s$ is \cite{buehler51,sear98}
\begin{equation}
\frac{a_s}{kT}=-\ln q_P
\label{adef}
\end{equation}
where $q_P$ is the partition function of a single particle
trapped in a cage formed by its twelve neighbours fixed at their
points in a perfect lattice. This
lattice is either face-centred cubic or hexagonal close-packed,
in each a particle has the same arrangement of nearest
neighbours so it does not matter which.
As the lattice constant
is sufficiently small, throughout the cage the particle attracts all
twelve neighbours and so its energy is constant at $-6\epsilon$.
The partition function is then just the volume available to the
centre of mass of the particle times $\Lambda^{-1}\exp(6\epsilon/kT)$,
where $\Lambda^{-1}$ is the integral over the momentum degrees of freedom.
The volume
is of the order of $(c\delta\sigma)^3$, i.e., the cube of the difference
between the lattice constant and the hard-sphere diameter.
The particle at its lattice position is $\sigma(1+c\delta)$
away from each of its neighbours and cannot move within $\sigma$
of them due to the hard-sphere interaction. Thus, it can move
a distance $c\delta\sigma$ in the direction of each of its neighbours.
Actually, the volume to which the particle is restricted is quite
complicated, see Ref. \citen{buehler51}, but it is not far
from $(c\delta\sigma)^3$.
We will approximate it by $(\delta\sigma)^3$; an approximation
which becomes increasingly accurate as $\delta$ decreases \cite{sear98}.
Then, we have for $q_P$,
\begin{equation}
q_P=(\delta\sigma)^3\Lambda^{-1}\exp(6\epsilon/kT)
\end{equation}
Inserting this expression for $q_P$ into Eq. (\ref{adef}),
\begin{equation}
\frac{a_s}{kT}=-\ln\left(\delta^3\right)
-6\frac{\epsilon}{kT}=
\frac{\mu_s}{kT}
\label{mus}
\end{equation}
where we have neglected a term $\ln(\Lambda/\sigma^3)$.
The chemical potential $\mu_s=a_s+p_s/\rho$ where $p_s$ is the
pressure and $\rho$ is the density.
At coexistence the pressure is very low,
note that the gas density is zero in Fig. \ref{sspd}.
Therefore, $p_s/\rho$
contributes a negligible amount to the chemical potential
for the $\beta_0$ model.
This enables us to equate $a_s$ and $\mu_s$
as we have done in Eq. (\ref{mus}).
For the $\beta_0$ model all temperatures below $T_{coll}$
in the $\delta\rightarrow0$ limit are low in the sense
that the solid is close-packed and so no different
from the solid at zero temperature.
When $\delta$ is nonzero the accuracy of
our low temperature approximation, Eq. (\ref{mus}), will rapidly worsen
as $T_{coll}$ is approached from below.

Equating the chemical potentials of the fluid, Eq. (\ref{muf}),
and solid, Eq. (\ref{mus}), phases yields an equation relating
the density of the coexisting fluid to the temperature
of coexistence, at a given value for $\delta$. We can rearrange the
equation to give the temperature of coexistence, $T_{co}$, as
a function of the density of the coexisting fluid phase, $\rho_f$,
\begin{equation}
\frac{kT_{co}}{\epsilon}
= \frac{6}{3\ln(1/\delta)-\ln\rho_f}.
\label{tco}
\end{equation}
As we cool a fluid phase of density $\rho_f$, then, at equilibrium,
the solid phase appears at the temperature $T_{co}$ of Eq. (\ref{tco}).
Using Eq. (\ref{tcoll}) for $T_{coll}$ we can express
$T_{co}$ as a fraction of $T_{coll}$
\begin{equation}
\frac{T_{co}}{T_{coll}}
=\frac{1}{1-\ln\rho_f/[3\ln(1/\delta)]}.
\label{tcotcoll}
\end{equation}
As $\delta$ is small the ratio $T_{co}/T_{coll}$ is near
one for a very wide range of $\rho_f$. For example, when
$\delta=0.05$, $T_{co}/T_{coll}=0.80$ and 0.66 for $\rho_f=0.1$ and
0.01, respectively.

\subsection{The second virial coefficient}

For the square-well potential of Eq. (\ref{monoss})
with $\delta\ll 1$, the second virial coefficient is
\begin{equation}
B_2= B_2^{hs}- 2\pi\sigma^3\delta(\exp(\epsilon/kT)-1),
\label{b2}
\end{equation}
where $B_2^{hs}=(2\pi/3)\sigma^3$
is the second virial coefficient of hard spheres.
Generally we will be working at sufficiently large
values of $\epsilon/kT$ that we can neglect the $-1$ term in
parentheses in Eq. (\ref{b2}). Therefore, we will neglect this term
from now on.
From the second virial coefficient Baxter defined the $\tau$
parameter \cite{baxter68}
\begin{equation}
\tau=\frac{1}{4(1-B_2/B_2^{hs})},
\end{equation}
which has been described as an effective temperature although it is not
a simple reduced temperature as it is not a simple ratio of the
thermal energy, $kT$, to an energy. It can be written directly in terms
of the temperature and the range $\delta$
\begin{equation}
\tau=\frac{1}{12\delta\exp(\epsilon/kT)}.
\label{tau}
\end{equation}
Straightaway, we can find $\tau$ at the temperature $T_{coll}$, $\tau_{coll}$,
by substituting Eq. (\ref{tcoll}) into Eq. (\ref{tau}),
\begin{equation}
\tau_{coll}=\frac{1}{12\delta^{1/2}},
\end{equation}
thus as $\delta$ tends to zero, $\tau_{coll}$ diverges.
$\tau_{coll}$ is the value of $\tau$ at $T_{coll}$, when
the fluid--solid transition widens from the
relatively narrow coexistence region of hard spheres to coexistence
between a dilute gas and a near close-packed solid.

The experiments measure the strength of the attractive interaction by
measuring the second virial coefficient via scattering whereas
we measure it via the ratio of the well-depth to the
thermal energy, $\epsilon/kT$. Thus we consider here the problem
of relating $B_2$ and $\tau$ to $\epsilon/kT$.
For nonzero $\delta$, $B_2$ and hence $\tau$  is a well-behaved
single-valued function of temperature and so it is possible to
uniquely map each value of $B_2$ or $\tau$ to a temperature $T$ and vice-versa.
However, if $\delta=0$ then $B_2$ is not a well-behaved function of
temperature. If we take the double limit $\delta$,
$kT/\epsilon\rightarrow0$ then \cite{stell91}
\begin{equation}
\frac{B_2}{B_2^{hs}}=
\left\{
\begin{array}{ll}
1& ~~~~~~ T > T_{B20}\\
-\infty & ~~~~~~ T < T_{B20}\\
\end{array}\right. ,
\end{equation}
where $T_{B20}$ is given by
\begin{equation}
\frac{kT_{B20}}{\epsilon}=\frac{1}{\ln(1/\delta)}.
\label{tb20}
\end{equation}
All temperatures above
$T_{B20}$ map onto a second virial coefficient equal to $B_2^{hs}$,
and all values below $T_{B20}$ map onto $-\infty$. Conversely,
{\it all} nonzero and non-infinite
values of $(B_2-B_2^{hs})/B_2^{hs}$,
and hence of $\tau$, correspond to the temperature
$T_{B20}$, see the discussion of Stell \cite{stell91}.
Which is why $\tau$ is useless to describe the phase diagram of
spheres with an attraction of zero range.

\begin{figure}
\caption{
Baxter's $\tau$ parameter, Eq. (\ref{tau}), plotted as a function
of the ratio of the thermal energy to the well depth, $kT/\epsilon$.
The solid, dashed, and dotted curves are for ranges
$\delta=0.1$, 0.05 and 0.01, respectively.
$kT_{B20}/\epsilon=0.43$ and 0.22 for $\delta=0.1$ and 0.01,
respectively.
}
\label{taufig}
\begin{center}
\epsfig{file=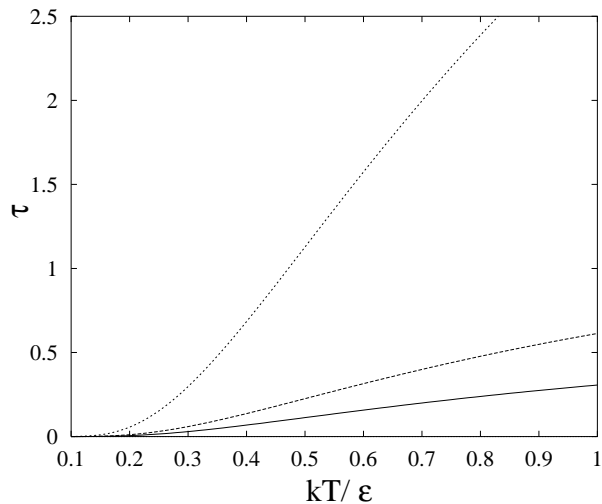,width=3.5in}
\end{center}
\end{figure}

Away from the double limit, where $\delta$ is small but
nonzero, the variation of $\tau$
with temperature depends strongly on the range $\delta$.
We plot $\tau$, Eq. (\ref{tau}), against
the reduced temperature $kT/\epsilon$ in Fig. \ref{taufig}.
At any given reduced temperature, $kT/\epsilon$, $\tau$ varies
as $\delta^{-1}$. Determining $\epsilon$ from scattering data
for $\tau$ thus requires an accurate knowledge of $\delta$.
This is not available for globular proteins, see section 4.

\subsection{The metastable fluid phase}

Nucleation occurs from a metastable phase, i.e., a phase
which is not the equilibrium phase. We cannot obtain the nucleation
rate in a metastable phase without making assumptions about this phase.
We will assume that the metastable phase is close to an ideal gas.
Justification for this assumption is provided by the fact that the densities
we will consider will be quite low $\rho\le0.1$ and the second
virial coefficient, Eq. (\ref{b2}), will not be too large,
$B_2/B_2^{hs}$ of order unity. The second virial coefficient is
the coefficient of the leading order correction to the ideal
gas pressure in a density expansion for the pressure \cite{hansen86}.
However, assuming ideal gas behaviour will not be correct if
the attractive interactions are strong enough to cause non-equilibrium
effects such as gelation or in the vicinity of a metastable
fluid--fluid transition \cite{tenwolde97}.
Both gelation \cite{ilett95,poon95}
and a metastable fluid--fluid transition \cite{broide91} have
been observed in experiment.

\section{Classical nucleation theory}

Classical nucleation theory
\cite{debenedetti,frenkel,laaksonen95,chaikin95}
assumes that the nucleus of the
solid phase has a free energy $\Delta F$ which is the sum
of a bulk term and surface term.
See the book of Debenedetti \cite{debenedetti} for an excellent introduction to
classical nucleation theory.
The bulk term is equal to the number of spheres in the nucleus, $n$,
times the chemical potential difference $\Delta\mu=\mu_{eq}-\mu_{m}$,
where $\mu_{eq}$ is the chemical potential of the coexisting equilibrium
fluid and solid phases, and $\mu_m$ is the chemical potential
of the metastable fluid phase which contains the nucleus.
Our Eq. (\ref{mus}) gives the chemical potential of the
solid phase at coexistence $\mu_s$ which is equal to $\mu_{eq}$.
The surface term is the surface area of the nucleus
times the surface tension $\gamma$ of the bulk
interface between the coexisting solid and fluid phases, i.e.,
any difference between the spherical interface between the solid nucleus
and the surrounding metastable fluid phase, and the flat interface between
the coexisting solid and fluid phases is ignored.
The interfacial tension of an interface between a
solid phase and a fluid phase depends
on the orientation of the interface with respect to the lattice of
the solid. To get around this problem,
either some orientationally averaged
interfacial tension or the interfacial tension with the interface
in the plane of a low index lattice plane are  used for $\gamma$.
The surface area of the nucleus is obtained by assuming that the
nucleus is a perfect sphere of radius $r$ which is related to $n$
by assuming that the density of spheres within the nucleus
is equal to the bulk density of the solid phase. For the
$\beta_0$ model this is the close-packed density of hard spheres
$2^{1/2}\sigma^{-3}$.
When $\delta$ is nonzero the
density is \cite{sear98} $2^{1/2}(1+c\delta)^{-3}\sigma^{-3}$,
where $c$ is a constant which is less than one \cite{sear98}.
We will see below that its value if irrelevant and so we do not
attempt to estimate it.
Then the surface area of the nucleus is
$n^{2/3}(18\pi)^{1/3}(1+c\delta)^2\sigma^2$.
So,
\begin{equation}
\Delta F(n) = n\Delta\mu + n^{2/3}(18\pi)^{1/3}\gamma(1+c\delta)^2\sigma^2
\label{df}
\end{equation}
The first term in $\Delta F$ is the bulk term,
which is negative and decreases linearly with $n$, and the
second term is the surface term
which is positive and increases as $n^{2/3}$. Thus
$\Delta F$ passes through
a maximum, denoted by $\Delta F^*$, at some value of $n$, 
denoted by $n^*$. The nucleus with $n^*$ spheres is called
the critical nucleus. This is the nucleus of the solid phase at
the size at which its free energy is greatest. As the
frequency with which a fluctuation occurs scales as
$\exp(-{\rm free~energy~cost}
/kT)$, the formation of the fluctuation with the highest free energy
cost is the rate limiting process. Thus, classical nucleation theory
predicts that the nucleation rate is proportional to
$\exp(-\Delta F^*/kT)$ where  $\Delta F^*=\Delta F(n^*)$ is the
maximum in the free energy of Eq. (\ref{df}).
Taking the derivative of Eq. (\ref{df}) with respect to $n$ and
equating to zero yields the number of spheres in the cluster at 
the top of the barrier
\begin{equation}
n^* = \frac{16\pi\left(\gamma(1+c\delta)^2
\sigma^2\right)^3}{3|\Delta\mu |^3},
\label{ncnt}
\end{equation}
and inserting this value of $n$ in Eq. (\ref{df})
\begin{equation}
\Delta F^* = \frac{8\pi\left(\gamma(1+c\delta)^2\sigma^2\right)^3}
{3\Delta\mu^2}.
\label{cnt}
\end{equation}

Now, the interfacial tension, $\gamma_{111}$, for an interface in the
111 crystal plane, between the coexisting close-packed solid phase
and infinitely dilute
gas phase of the $\beta_0$ model is given by \cite{sear99}
\begin{equation}
\gamma_{111}=3^{1/2}\epsilon\sigma^{-2}~~~~~\delta=0.
\end{equation}
Which, like our expression for the chemical potential, Eq. (\ref{mus}),
is a low temperature approximation also valid when $0<\delta\ll1$,
but which breaks down, when $\delta$ is nonzero, as
$T_{coll}$ is approached from below.
Other lattice planes have slightly higher tensions,
e.g., the 100 plane has a surface tension
$\gamma_{100}=2\epsilon\sigma^{-2}$. We will approximate
$\gamma$ for all small values of $\delta$ by
the surface tension of the 111 plane $\gamma_{111}$, which will
vary with $\delta$.
The tension $\gamma_{111}$ is the energy penalty due to the broken
bonds at the surface of the solid \cite{sear99}. Therefore,
it scales as one over the square of the lattice constant,
$(1+c\delta)\sigma$.
So, for all small values of $\delta$ we have
\begin{equation}
\frac{\gamma(1+c\delta)^2\sigma^2}{kT}
=3^{1/2}\frac{\epsilon}{kT}=
\frac{3^{1/2}}{2}\ln(1/\delta)\left(\frac{T_{coll}}{T}\right),
\label{gamma}
\end{equation}
where the second equality was obtained using Eq. (\ref{tcoll}).
In Ref. \citen{sear99} it was shown that the gas-solid interface is
narrow, only one sphere wide, and so the assumption of
classical nucleation theory that the interface is narrow
\cite{frenkel,laaksonen95,debenedetti} is fully justified.

Then, the free-energy barrier to nucleation, $\Delta F^*$,
expressed as a function of temperature and the excess
chemical potential over that at equilibrium, is
\begin{equation}
\frac{\Delta F^*}{kT} =
\frac{3^{1/2}\pi\left[\ln(1/\delta)\right]^3 (T/T_{coll})^{-3}}
{(\Delta\mu/kT)^2}.
\label{cntmu}
\end{equation}
The corresponding expression for $n^*$ is
\begin{equation}
n^* =
\frac{3^{1/2}2\pi\left[\ln(1/\delta)\right]^3 (T/T_{coll})^{-3}}
{|\Delta\mu/kT|^3}.
\label{nstar}
\end{equation}
Equations (\ref{cntmu}) and (\ref{nstar}) are valid below and not too
close to $T_{coll}$. From them we see that at fixed
chemical potential difference, $\Delta\mu/kT$, and
ratio of the temperature to $T_{coll}$,
as $\delta$ decreases
both the free-energy barrier and the size of the
critical nucleus increase as $[\ln(1/\delta)]^3$.
The increasing surface tension, Eq. (\ref{gamma}),
as $\delta$ decreases is increasing the barrier to nucleation.

We would like to express $\Delta F^*$ and $n^*$ purely in terms
of temperature, density of the metastable fluid phase, $\rho_f$,
and $\delta$. To do this we need to know
$\Delta\mu=\mu_s-\mu_m$. Now, $\mu_s$ is given by Eq. (\ref{mus}).
This leaves the effective chemical potential of the metastable fluid
phase. We will estimate this by {\it assuming} that the metastable fluid
phase closely resembles an ideal gas, see Section 2.3.
This is an accurate approximation near coexistence.
So, with this assumption, $\mu_m$ is given by Eq.
(\ref{muf}) with $\rho=\rho_f$, and
so $\Delta\mu$ is
\begin{equation}
\frac{\Delta\mu}{kT}=
3\ln(1/\delta)-6\frac{\epsilon}{kT}-\ln\rho_f
\label{deltamu1}
\end{equation}
If we use Eq. (\ref{tco}) to obtain
a value for $T_{co}$ at the given value for $\rho_f$
then we can define a reduced temperature
$t=T/T_{co}$; a function of $T$ and $\rho_f$.
The reduced temperature $t$ is the temperature as a fraction of the
coexistence temperature at the same value of the fluid density.
In terms of $t$, Eq. (\ref{deltamu1}) can be rewritten as
\begin{equation}
\frac{\Delta\mu}{kT}=
3\ln(1/\delta)[1-t^{-1}]\left[1-\frac{\ln\rho_f}{3\ln(1/\delta)}\right].
\label{deltamu}
\end{equation}
This is the excess chemical potential of a metastable
fluid phase at a density $\rho_f$ and at a fraction $t$ of the
temperature at which a fluid of this same density $\rho_f$ coexists
with the solid phase.
$\Delta\mu$ is of order $kT$ when $(T_{co}-T)/T_{co}=1-t
\sim 1/\ln(1/\delta)$. To achieve a chemical potential in the
metastable fluid which is of order $kT$ higher than at
equilibrium we need only go a small fraction of $T_{co}$
below $T_{co}$.
When the temperature difference $(T_{co}-T)/T_{co}$ is finite,
$\Delta\mu$ diverges in the $\delta\rightarrow0$ limit.

We can substitute Eq. (\ref{deltamu}) for $\Delta\mu$,
and Eq. (\ref{tcotcoll}) for $T_{coll}$ into
Eq. (\ref{cntmu}) to obtain the free energy barrier to
nucleation as a function of temperature and the range
of the attraction, $\delta$,
\begin{equation}
\frac{\Delta F^*}{kT} =
\frac{3^{-3/2}\pi\ln(1/\delta)
\left(1-\ln\rho_f/[3\ln(1/\delta)]\right)}
{t(1-t)^2}.
\label{cntt}
\end{equation}
Which is the free energy barrier to the nucleation of the solid phase
in a metastable
fluid phase at a density $\rho_f$ and at a fraction $t$ of the
temperature at which a fluid of this same density $\rho_f$ coexists
with the solid phase.
At a given fraction, $t$, of the phase transition
temperature, the free energy barrier to nucleation increases as
$\ln(1/\delta)$, as the range, $\delta$, decreases.
The corresponding expression for $n^*$ is very simple,
\begin{equation}
n^* =
\frac{3^{-5/2}2\pi }
{(1-t)^3},
\label{nt}
\end{equation}
which is not a function
of the range $\delta$ at fixed $t$.
It is the number of spheres in the critical nucleus in a metastable
fluid phase at a density $\rho_f$ and at a fraction $t$ of the
temperature at which a fluid of this same density $\rho_f$ coexists
with the solid phase.
Our Eqs. (\ref{cntt}) and (\ref{nt})
for $\Delta F^*$ and $n^*$, respectively, rely on our approximations for
$\gamma$ and $\Delta\mu$ being accurate, which they are provided
we are not too close to $T_{coll}$ when $\delta$ is nonzero,
and provided the metastable phase is not too unlike an ideal gas.
Also, the free energy is derived by assuming that we can treat it
as the sum of a bulk part and an interfacial part. This is reasonable
if $n$ is large but breaks down when it is small. Thus,
classical nucleation theories break down when $n^*$ drops below
about twenty. From Eq. (\ref{nt}), $n^*=20$ when $t=0.73$.
Thus below $t\simeq0.7$ the critical nucleus is so small that
classical nucleation theory can no longer be expected to be reliable.

\begin{figure}
\caption{
The free energy barrier $\Delta F^*/kT$ is plotted as a function
of $t=T/T_{co}$ for three ranges:
$\delta=0.1$ (solid curve), $\delta=0.05$ (dashed curve) and
$\delta=0.01$ (dotted curve). For $\delta=0.1$, 0.05 and 0.01,
$kT_{co}/\epsilon=0.65$, 0.53 and 0.37, respectively.
}
\label{ftfig}
\begin{center}
\epsfig{file=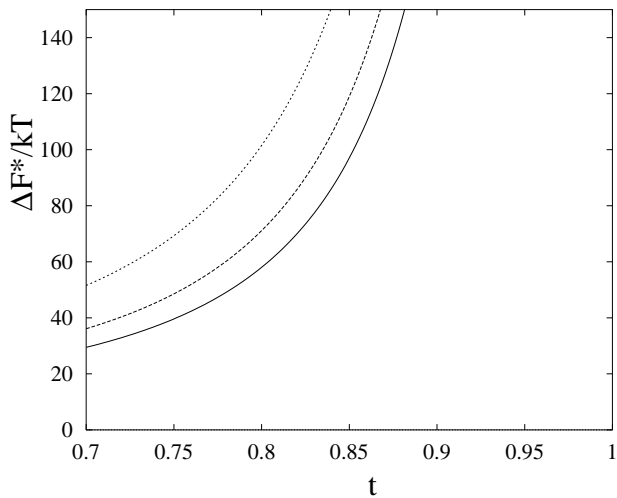,width=3.5in}
\end{center}
\end{figure}

\begin{figure}
\caption{
The number of spheres at the top of barrier, $n^*$, as a function
of $t=T/T_{co}$. The one curve plotted is for any (short) range
$\delta$.
}
\label{ntfig}
\begin{center}
\epsfig{file=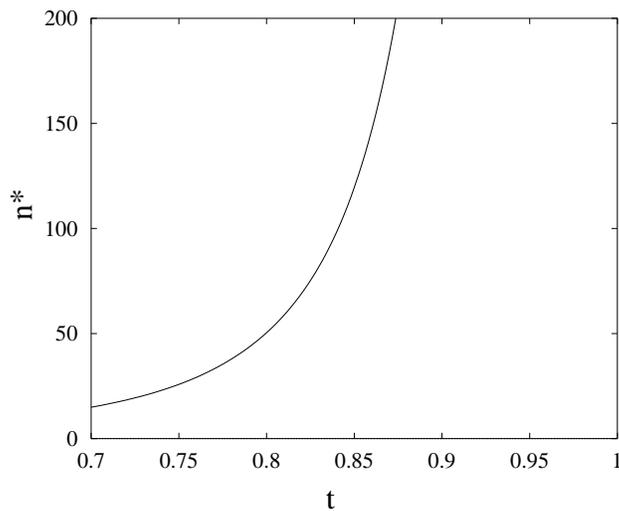,width=3.5in}
\end{center}
\end{figure}

The free energy barrier $\Delta F^*/kT$ and the number
of spheres $n^*$ at the top of the barrier are plotted in Figs.
\ref{ftfig} and \ref{ntfig}, respectively, for three
different ranges $\delta=0.1$, 0.05 and 0.01. 
At any value of $t$ the shorter the
range the larger is $\Delta F^*/kT$. This is true even though as the
range decreases $\Delta\mu/kT$, Eq. (\ref{deltamu}), increases.
$\Delta F^*/kT$ varies as $(\gamma\sigma^2/kT)^3/(\Delta\mu/kT)^2$ and as both
the surface tension $\gamma$ and $\Delta\mu/kT$ vary as
$\ln(1/\delta)$, then $\Delta F^*/kT$ varies as $\ln(1/\delta)$,
Eq. (\ref{cntt}). But the number of spheres at the top
of the barrier is independent of range,
see Fig. \ref{ntfig}. Thus
we predict that when the attraction is very short-ranged,
the free energy barrier to nucleation is larger than
it is when the attraction is longer-ranged but
the number of spheres at the top of the free energy barrier, $n^*$,
is the same.
As both the free-energy barrier and $\Delta\mu/kT$ increase
with decreasing range then the free-energy barrier will only
be small enough for nucleation to occur when $\Delta\mu/kT$ is very
large.

\section{Comparison with protein nucleation}

We have specified the state of our model system by the density $\rho$
and the reduced temperature $kT/\epsilon$, which is
the ratio of the thermal energy to the
attractive energy. For protein solutions the
attraction is a free energy not an energy, and
so its strength $\epsilon$ is a function of temperature,
salt concentration and other variables. The
dependence on temperature is then rather complex.
For example, adding polymer induces a depletion attraction
\cite{asakura,vrij76} which is proportional to temperature and
to the concentration of polymer. All this makes a direct
determination of the ratio of the thermal energy to the
attraction free energy all but impossible in experiment.
What can often be measured in experiment is the second
viral coefficient. George and Wilson \cite{george94}
use the second virial
coefficient $B_2$ as a measure of the strength of the attraction, and
Rosenbaum {\it et al.} \cite{rosenbaum96} use Baxter's $\tau$
parameter. The second virial coefficient, Eq. (\ref{b2}),
is a function of $\epsilon/kT$ and $\delta$.
Thus to convert a temperature into $\tau$ or vice versa we need to know
the range of the attraction between protein molecules. We do not know
this range,
also we do not know the extent to which it varies from protein
to protein. Rosenbaum {\it et al.} \cite{rosenbaum96} study
lysozyme, effective diameter 3.4nm and STA, effective diameter
1.3nm. Only if the range of the attraction scales with the size
of the protein will the range of the attraction be the same in each case.
In the absence of an estimate for the range of the attraction between
protein molecules we have performed calculations for a
few values of the range, $\delta$.

\begin{figure}
\caption{
The free energy barrier $\Delta F^*/kT$ is plotted as a function
of Baxter's $\tau$ parameter for three ranges:
$\delta=0.1$ (solid curve), $\delta=0.05$ (dashed curve) and
$\delta=0.01$ (dotted curve). All three curves are plotted down
to $\tau=0.1$ but in the case of $\delta=0.01$ the size of
critical nucleus, $n^*$, drops below twenty at $\tau=0.2$ and so
the dotted curve from $\tau=0.1$ to 0.2 is really out
of the range of applicability of classical nucleation theory.
}
\label{ftaufig}
\begin{center}
\epsfig{file=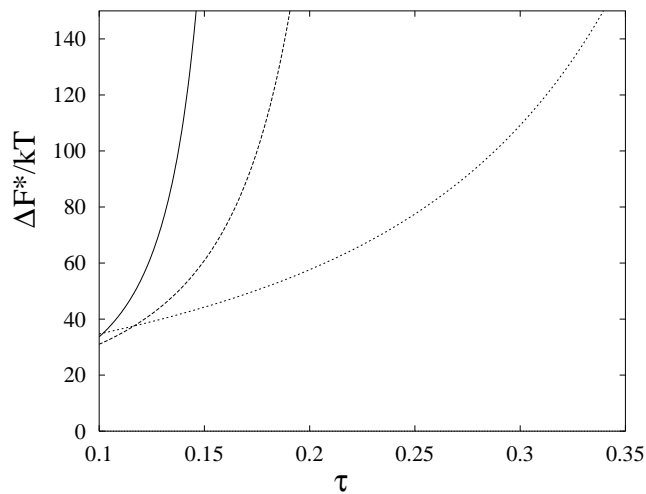,width=3.5in}
\end{center}
\end{figure}

Rosenbaum {\it et al.} \cite{rosenbaum96} plot
the values of Baxter's $\tau$ parameter for which a number
of globular proteins crystallise.
They conclude that \lq\lq crystallising conditions occur for
$0.06 < \tau <0.15$".
In order to compare the results of classical nucleation theory
with experiment we rewrite the free-energy barrier, Eq. (\ref{cnt}),
in terms of $\tau$. To do this we use Eq. (\ref{tau}) to substitute
$\tau$ for $\epsilon/kT$ in Eqs. (\ref{gamma}) and (\ref{deltamu1})
and then we substitute the resulting expressions for
$\gamma$ and $\Delta\mu$ into Eq. (\ref{cnt}) to obtain
\begin{equation}
\frac{\Delta F^*}{kT}=
\frac{ 3^{1/2}8\pi|\ln12\delta\tau|^3}
{(3\ln(1/\delta)-\ln\rho_f+6\ln12\delta\tau)^2}.
\end{equation}
We have plotted $\Delta F^*/kT$ as a function of $\tau$
in Fig. \ref{ftaufig}.

For $\delta=0.1$ and 0.05, the free-energy barrier is a very sharply
varying function of $\tau$. A little below $\tau=0.15$ and 0.2,
for $\delta=0.1$ and 0.05, respectively, the free-energy barrier
increases to huge values.
The sharply varying
free energies and the fact that the nucleation rate depends
exponentially on the free-energy means that there will be a well defined
upper limit to the value of $\tau$ at which nucleation occurs
during a given time scale.
Our prediction, for very reasonable ranges $\delta=0.05$ and 0.1,
of an upper limit to nucleation at $\tau$ around 0.15 is in good agreement
with experiment. Despite the crudity of our theory this
agreement may not be fortuitous.
We do not know the range $\delta$ that best corresponds to
the interaction potential of
any globular protein and our calculation of the free energy barrier
may be out by tens of $kT$. However, when the range is
of the order of a few \% of the hard-sphere diameter then
changing the range by a factor of two or more and adding or
subtracting several tens of $kT$ to $\Delta F^*$ do not alter our
prediction that nucleation cannot occur above $\tau=0.15\pm0.05$.
Because of the nucleation rate's exponential dependence on the
distance into the metastable regime
it is common to observe a well-defined threshold beyond which
nucleation is not observed \cite{debenedetti,laaksonen95}.
Thus our finding of a well-defined upper limit on nucleation is
no surprise. However, the fact that this upper limit occurs at a value
of $\tau$ which depends very weakly on the range of the attraction,
$\delta$, is more interesting. It is a possible explanation of why
many different proteins have approximately the same upper limit
even though the effective interaction
potential may differ from protein to protein.

If we compare Figs. \ref{ftfig} and \ref{ftaufig} we see that when
the barrier height is plotted as a function of $t$, the shorter the range
the steeper the curve followed by the barrier height, whereas when
the barrier height is plotted as a function of $\tau$ it is the other
way around. Changing variables from $t$ to $\tau$ reverses
the trend in the change of the free-energy barrier with $\delta$.
This is a consequence of the increasingly
rapid variation of $\tau$ with temperature as $\delta$ decreases.
It suggests caution in
interpreting $\tau$ as a temperature-like variable.

Although the interaction between globular proteins
is complex and poorly understood,
there is an experimental system with a known, short range
\cite{ilett95,poon95}. This
is hard-sphere-like colloidal particles with added non-adsorbing
polymer with a radius of gyration much less than the diameter
of the particles. Then the polymer induces a depletion attraction
\cite{asakura,vrij76} with a range $\delta$ approximately equal
to the radius of gyration divided by the diameter of
the colloidal particles. Ilett {\it et al.} \cite{ilett95,poon95}
studied a system with a range $\delta$ of approximately 0.08.
However, as the coexistence densities were not determined
accurately we cannot compare our theoretical predictions with
these experiments. Qualitatively,
they find that crystallisation only occurs if they
are not too far into the metastable regime, if $t$ is not too small.
If $t$ is too small then a gel forms. As for proteins there
is a window within which crystallisation takes place.

\section{Conclusion}

We have calculated the free-energy barrier to nucleation, within classical
nucleation theory, for spherically symmetric potentials
consisting of a hard-core and a short-range attraction. A simple model
potential for globular proteins and some colloidal systems.
The short-range attraction makes nucleation more difficult:
the surface tension, Eq. (\ref{gamma}), is larger than for many other systems,
e.g., hard spheres, which means that at the same value of $\Delta\mu/kT$
the free energy barrier to nucleation is higher when there is a short-ranged
attraction than for hard spheres. The surface tension for hard spheres is,
from density-functional theory, $0.3kT\sigma^{-2}$
\cite{ohnesorge94,kyrlidis95},
whilst with an attraction of range 10\% of the diameter
of the particle, $\delta=0.1$, at a temperature $(2/3)T_{coll}$,
it is $5.9kT\sigma^{-2}$.
The surface tension for hard spheres is an order
of magnitude smaller and the nucleation rate is very sensitive
to the surface tension, it varies as $\exp[-(\gamma\sigma^2/kT)^3]$
at constant $\Delta\mu/kT$.

The large barriers to nucleation predicted by classical nucleation
theory are certainly consistent with the difficulty experienced
in crystallising proteins \cite{durbin96,rosenberger96,george94,rosenbaum96}
and the non-equilibrium behaviour observed for colloidal particles
with a short-ranged attraction \cite{ilett95,poon95}.
However, the uncertainty surrounding the interaction between protein
molecules makes comparison with theory, which is for a specific
model potential, very difficult.
Experimental results for the location of the onset of nucleation
\cite{debenedetti,laaksonen95} for colloidal
systems for which the
the interaction potential potential and
phase coexistence boundaries are both known
would be very useful in understanding protein crystallisation and
in testing theoretical predictions.

The barrier to nucleation decreases as the temperature drops more and
more below the transition temperature $T_{co}$, i.e., as $t$ decreases,
and we move further into the metastable region of the
phase diagram. Thus we can lower the free energy barrier
by moving deeper into the metastable regime.
However, by doing so we may encounter difficulties such
as the fluid becoming sufficiently cold to form a gel-like
state \cite{ilett95,poon95,poon97} in which the dynamics are at least
partly arrested, preventing the formation of the crystalline phase.
Ten Wolde and Frenkel \cite{tenwolde97} have suggested that a
metastable fluid--fluid transition may offer a way round this
problem.

An interesting prediction of classical nucleation theory is that
the critical nucleus at a given value of the free-energy barrier
shrinks as the range of the attraction, $\delta$, decreases.
Thus under the conditions in which the free-energy barrier
is small enough to permit nucleation on a reasonable
time scale the nucleus becomes smaller and smaller as $\delta$ decreases.
Indeed it becomes so small that classical nucleation theory breaks
down, even when $\Delta F^*$ is still large, of order tens of $kT$.



\end{multicols}

\newpage
~~\\


~~\\


~~\\


~~\\


~~\\

\end{document}